\documentclass[journal, 10pt]{IEEEtran}  
	
\IEEEoverridecommandlockouts                             
\usepackage[mode=tex]{standalone} 

\usepackage[utf8]{inputenc}
\usepackage{lipsum} 
\usepackage{amsmath, amssymb} 
\usepackage{cite} 
\usepackage{amsthm}

\usepackage{bm}
\usepackage{graphicx}
\usepackage[font={footnotesize}]{caption}
\usepackage{xcolor}
\usepackage{subfig} 
\usepackage{booktabs}
\usepackage{multirow}
\usepackage{ulem} 
\usepackage{tikz,pgfplots}
\usepackage{url}

\definecolor{DarkGreen}{rgb}{0.0, 0.5, 0.0}

\newcommand{\Removed}[1]{%
}%

\newcommand{\RemovedMath}[1]{%
    \ifmmode\text{\sout{\ensuremath{#1}}}\else\sout{#1}\fi%
}%

\newcommand{\NoRev}[1]{%
	{
	#1%
	}%
}%

\newcommand{\RevA}[1]{%
	{
	#1%
	}%
}%

\newcommand{\RevB}[1]{%
	{
	#1%
	}%
}%

\newcommand{\MinRevB}[1]{%
	{
	#1%
	}%
}%

\newcommand{\RevC}[1]{%
	{
	#1%
	}%
}%

\newcommand{\MinRevC}[1]{%
	{
	#1%
	}%
}%

\newcommand{\RevD}[1]{%
	{
	#1%
	}%
}%

\newcommand\xqed[1]{%
\leavevmode\unskip\penalty9999 \hbox{}\nobreak\hfill
\quad\hbox{#1}}
\newcommand\demo{\xqed{$\triangle$}}

\usepackage[%
	ruled,%
	vlined,%
	boxed,%
	linesnumbered,%
	commentsnumbered]{algorithm2e}%
	
\usepackage{etoolbox}
\makeatletter
\patchcmd{\@algocf@start}
  {-1.5em}
  {0pt}
  {}{}
\makeatother

\pgfplotsset{compat=newest}%
\usetikzlibrary{positioning,matrix,shapes.multipart,shapes.misc,spy}%
\tikzstyle{annotation}=[fill=white]%

\newcommand{\vect}[1]{\boldsymbol{#1}}
\newcommand{\imag}{\jmath}

\newcommand{\SNR}{\text{SNR}}

\usetikzlibrary{shapes.arrows}
\usetikzlibrary{spy}

\tikzset{%
	partial ellipse/.style args={#1:#2:#3}{%
		insert path={+ (#1:#3) arc (#1:#2:#3)}%
	}%
}%

\tikzstyle{star marker}=[mark=star, mark options={solid, scale=1}]
\tikzstyle{diamond marker}=[mark=diamond*, mark options={solid, fill=white, mark size=2.0pt}]
\tikzstyle{triangle marker}=[mark=triangle*, mark options={solid, scale=0.8}]
\tikzstyle{square marker}=[mark=square*, mark options={solid, scale=0.8}]
\tikzstyle{circle marker}=[mark=*, mark options={solid, scale=1, fill=white}]

\title{Data-Driven Estimation of \MinRevB{Capacity Upper} Bounds}
\author{%
Christian Häger, \emph{Member, IEEE} and 
Erik Agrell, \emph{Fellow, IEEE}
\thanks{This work was supported by the Knut and Alice Wallenberg Foundation (grant no.~2018.0090) and the Swedish Research Council (grants no.~2020-04718 and 2021-03709).
The computations were enabled by resources provided by the Swedish National Infrastructure for Computing (SNIC) at the Chalmers Centre for Computational Science and Engineering (C3SE) partially funded by the Swedish Research Council (grant no.~2018-05973).}%
\thanks{%
C.~H\"{a}ger and E.~Agrell are with the Department of Electrical Engineering, Chalmers University of Technology, 41296 Gothenburg, Sweden (email: \{hagerc,  agrell\}@chalmers.se).}%
\thanks{Manuscript received Month xx, 2021, revised Month xx, 2021}%
}%
\date{November 2021}

\begin{document}

\maketitle

\begin{abstract}
We consider the problem of estimating an upper bound on the capacity of a memoryless channel with unknown channel law and continuous output alphabet. 
A novel data-driven algorithm is proposed that exploits the dual representation of capacity where the maximization over the input distribution is replaced with a minimization over a reference distribution on the channel output. 
To efficiently compute the required divergence maximization between the conditional channel and the reference distribution, we use a modified mutual information neural estimator that takes the channel input as an additional parameter.
We \RevA{numerically} evaluate our approach on different memoryless channels and show \RevA{empirically} that the estimated upper bounds closely converge either to the channel capacity or to best-known lower bounds.

\end{abstract}

\section{Introduction}

The capacity of a communication channel is the maximum rate that can be reliably transmitted \cite{Shannon1948}. 
Even though capacity is of fundamental importance for both theory and practice, exact analytical expressions are only available in relatively few cases. 
If the underlying channel law is known, numerical techniques can be used to approximately compute capacity such as the well-known Blahut--Arimoto algorithm \cite{Blahut1972, Arimoto1972} and its many variations, see, e.g., \cite{Naiss2013} and references therein. 

Recently, there has been significant interest in developing capacity estimation algorithms based on machine learning\RevA{\Removed{, without assuming knowledge about the underlying channel law}} \cite{Shen2018ecoc, Fritschek2019, Aharoni2020, Fritschek2020, Mirkarimi2021, Letizia2021, Mirkarimi2021comparison}. 
These approaches have their origin in the seminal work \cite{OShea2017}, where the authors propose to reinterpret the communication problem as a reconstruction task using parameterized transmitters and receivers, similar to autoencoders (AEs) in machine learning. 
It can be shown that the cross-entropy minimization commonly used to train AEs maximizes a lower bound on mutual information, whereas the transmitter optimization can be regarded as shaping a discrete input distribution. 
Using this approach, tight lower bounds on the capacity of a nonlinear phase noise (NLPN) channel were for example obtained in \cite{Shen2018ecoc}. 

A disadvantage of the AE approach is that it requires a differentiable channel model to compute gradients for the transmitter optimization. 
\RevA{\Removed{Therefore, it cannot be used directly in practical settings where the channel is only accessible via input--output samples.}}
To address this problem, \cite{Fritschek2019} proposes to use the sample-based mutual information neural estimation (MINE) technique from \cite{Belghazi2018} and train the AE transmitter based on the (differentiable) MINE. 
Related approaches were subsequently proposed for \RevC{\Removed{feedback channels with memory} more general channels that may include feedback and/or memory}\footnote{See also the earlier work in \cite{Aharoni2019} based on reinforcement learning, which, however, requires knowledge about the channel law.} in \cite{Aharoni2020} and for memoryless multiple-access channels in \cite{Mirkarimi2021}. 
Moreover, a hybrid approach that regularizes the cross-entropy-based AE training using MINE is proposed in \cite{Letizia2021}. 
Comparisons of various sample-based mutual information estimators similar to MINE can be found in \cite{Fritschek2020} and \cite{Mirkarimi2021comparison}.

All of the above learning-based approaches target the estimation of \emph{lower} capacity bounds using the conventional maximization of mutual information (or directed information in \cite{Aharoni2020}) over the input distribution, see \eqref{eq:capacity} below. 
In this paper, we follow a different path and consider the problem of estimating an \emph{upper} capacity bound by exploiting the dual representation of channel capacity, which is described in Sec.~\ref{sec:dual}.    
Our work relies on a variation of MINE for estimating the divergence between the conditional channel and a given reference distribution. 
\RevA{\Removed{The resulting method  can be applied to memoryless channels with a continuous output alphabet. 
Similar to \mbox{\cite{Shen2018ecoc, Fritschek2019, Aharoni2020, Fritschek2020, Mirkarimi2021, Letizia2021, Mirkarimi2021comparison}}, it is agnostic to the underlying channel law.}}
\RevA{Compared to the conventional Blahut--Arimoto algorithm, the main advantage of our approach and similar works on neural capacity estimation in \cite{Shen2018ecoc, Fritschek2019, Aharoni2020, Fritschek2020, Mirkarimi2021, Letizia2021, Mirkarimi2021comparison} stems from the fact that no knowledge about the underlying channel law is required. 
As such, the resulting estimators can be used in settings where the channel is only accessible via input--output samples (e.g., in experimental setups) and the precise channel law is unknown. 
}

\emph{Notation:}
Random variables are denoted by upper-case letters (e.g., $X$), realizations by lower-case letters (e.g., $x$), and sets by calligraphic letters (e.g., $\mathcal{X}$). 
The probability distribution of a random variable $X$ is denoted by $f_X$. 
\MinRevC{Expectation is denoted by $\mathbb{E}[\cdot]$, mutual information by $I(\cdot; \cdot)$, and Kullback--Leibler divergence by $D(\cdot||\cdot)$.}
For an integer $N$, we define the set $[N] = \{1, 2, \ldots, N\}$.

\section{Dual Representation of Channel Capacity}
\label{sec:dual}

We consider a memoryless channel\MinRevC{\footnote{To keep the notation simple, we focus on scalar channels. 
However, our approach generalizes to (block-wise) memoryless channels where the input and outputs are (possibly complex-valued) random vectors, see Sec.~\ref{sec:nlpn}.}} with input $X \in \mathcal{X}$ and output $Y \in \mathcal{Y}$.
The channel law conditioned on a particular input $x$ is denoted by $f_{Y|X=x}(y)$. 
In general, the input is assumed to be constrained by a cost function $c : \mathcal{X} \to \mathbb{R}_{\geq 0}$. 
\RevD{\Removed{The application of $c$ to an input distribution $f_X$ is defined by $c(f_X) = \mathbb{E}_{f_X}[c(X)]$ \mbox{\cite[p.~108]{Csiszar1981}}.}}
The capacity of such a channel is 
\begin{align}
    \label{eq:capacity}
    C = \max_{f_X : \RevD{\mathbb{E}_{f_X}[c(X)]} \leq P} I(X; Y), 
\end{align}
where $P$ denotes the maximum average cost. 
In the following, we work with the dual representation \cite[p.~142]{Csiszar1981}
\begin{align}%
    \label{eq:cup2}
    C = \min_{\gamma \geq 0} \left[ F(\gamma) + \gamma P \right],%
\end{align}%
where
\begin{align}%
    \label{eq:cup}
    F(\gamma) = \min_{q_Y} \max_{x \in \mathcal{X}} \left[ D(f_{Y|X=x} || q_Y) - \gamma c(x) \right].%
\end{align}%
The minimization in \eqref{eq:cup} is over all distributions $q_Y$ on $\mathcal{Y}$. 
Note that any fixed choice for the reference distribution $q_Y$ leads to an upper bound in \eqref{eq:cup} and hence on the capacity \eqref{eq:cup2}. 

According to \cite{Thangaraj2017}, the above dual approach first originated in \cite{Topsoe1967} and was further developed in \cite{Kemperman1974}, \cite{Csiszar1981}, and \cite{Lapidoth2003}.
Recent work exploiting this approach has mostly focused on choosing $q_Y$ to obtain a tractable analytical expression for the resulting upper bound, see, e.g., \cite{Lapidoth2003, Thangaraj2017}. 
In this paper, we will instead use \eqref{eq:cup} as a blueprint for an iterative optimization procedure that alternates between training a divergence estimator (see Sec.~\ref{sec:divergence}) and the reference distribution $q_Y$ (see Sec.~\ref{sec:ndt}) based on the obtained estimator. The resulting algorithm is described in Sec.~\ref{sec:algorithm}. 

\section{Data-Driven Divergence Estimation}
\label{sec:divergence}

We use the approach proposed in \cite{Belghazi2018} to estimate the divergence term in \eqref{eq:cup}. 
This approach is based on the Donsker--Varadhan \RevC{(DV)} representation \cite[Th.~1]{Belghazi2018} 
\begin{align}
	\label{eq:donsker}
	D(f_{Y|X=x}||q_Y) = \sup_{T \in \mathcal{T}} \mathbb{E}_{f_{Y|X=x}}[T(Y)] - \log\MinRevC{\left( \mathbb{E}_{q_Y}[ e^{T({Y})} ] \right)},
\end{align}
where the supremum is over all functions $T \MinRevC{\colon} \mathcal{Y} \to \mathbb{R}$ such that the expectations in \eqref{eq:donsker} are finite. 
The idea in \cite{Belghazi2018} is to approximate the class of functions $\mathcal{T}$ using a neural network (NN) $T_\theta \MinRevC{\colon} \mathcal{Y} \to \mathbb{R}$, where $\theta$ are the NN parameters. 
$T_\theta$ is also referred to as the statistics network. 
For a fixed set of parameters $\theta$, the resulting estimator is 
\begin{align}
	\label{eq:mine_objective}
	\hat{D}_\theta = \frac{1}{B} \sum_{i=1}^B T_\theta(y^{(i)}) - \log\left( \frac{1}{B} \sum_{i=1}^B e^{T_\theta(\tilde{y}^{(i)})} \right),
\end{align}
where $B$ is the batch size, $y^{(1)}, \ldots, y^{(B)} \sim f_{Y|X=x} $, and $\tilde{y}^{(i)}, \ldots, \tilde{y}^{(B)} \sim q_Y$. 
This estimator can be iteratively trained by running gradient ascent on \eqref{eq:mine_objective}, see \cite[Alg.~1]{Belghazi2018} for details. 

\smallskip
\emph{Remark:}
We use the above estimator mainly for its simplicity and the fact that it empirically tends to work well (see, e.g., \cite{Mirkarimi2021comparison}). 
\RevB{However, one should be aware that using the lower-bound estimator \eqref{eq:mine_objective} does not guarantee to result in a true upper bound on capacity, which would in principle require a full optimization over $\mathcal{T}$.}
\RevB{Moreover,}
the gradients resulting from \eqref{eq:mine_objective} are biased \cite{Belghazi2018} \RevB{and} the estimator has high variance, especially if the true underlying divergence is large \cite{McAllester2020}. \NoRev{\Removed{, and it provides neither a guaranteed upper nor lower bound on divergence for a finite sample size \mbox{\cite{Poole2019}}.}}
We comment on potential alternative approaches in Sec.~\ref{sec:conclusion}.
\smallskip

Note that the estimator \eqref{eq:mine_objective} assumes a fixed channel input $x$. 
A different statistics network $T_\theta$ would thus be required for each input to evaluate the maximization in \eqref{eq:cup} over the input alphabet. 
However, this quickly becomes infeasible if the size of the input alphabet is large or infinite. 
To circumvent this problem, we propose a modified version of \eqref{eq:mine_objective} where the input $x$ is taken as an additional input to the statistics network $T_\theta$, i.e., $T_\theta \MinRevC{\colon} \mathcal{Y} \times \mathcal{X} \to \mathbb{R}$. 
The resulting modified estimator is denoted by 
\begin{align}
    \begin{split}
	\label{eq:modified_mine}
	\hat{D}_\theta(x)  
  = \frac{1}{B} \sum_{i=1}^{B} T_\theta(y^{(i)}, x)  
  - \log\left( \frac{1}{B} \sum_{i=1}^{B} e^{T_\theta(\tilde{y}^{(i)}, x)} \right).
  \end{split}
\end{align}
This formulation allows us to train a single statistics network that works well for a range of different channel inputs. 
In particular, this can be accomplished by jointly considering multiple inputs $\{x^{(1)}, \ldots, x^{(N_t)}\} = \mathcal{X}_t \subseteq \mathcal{X}$ and running gradient descent on an average loss \RevC{$-\frac{1}{N_t} \sum_{j} \hat{D}_\theta(x^{(j)})$}. \RevC{\Removed{defined by}}
\RevC{However, averaging the logarithms in \eqref{eq:modified_mine} leads to a geometric mean 
$\frac{1}{N_t} \sum_{j} \log a_j = \log ((\prod_{j} a_j )^{1/{N_t}})$. 
We found that replacing this geometric mean with an arithmetic mean gives a numerically more stable training behavior, resulting in 
\MinRevC{\begin{align}
    \label{eq:modified_mine_objective}
     L_\theta = &- \frac{1}{N_t B} \sum_{j = 1}^{N_t} \sum_{i=1}^{B} T_\theta(y^{(i,j)}, x^{(j)})  \nonumber\\
  &+ \log\left( \frac{1}{N_t B} \sum_{j=1}^{N_t}\sum_{i=1}^{B} e^{T_\theta(\tilde{y}^{(i,j)}, x^{(j)})} \right),   
\end{align}}
where $y^{(1,j)}, \ldots, y^{(B,j)} \sim f_{Y|X=x^{(j)}}$
and
$\tilde{y}^{(1,j)}, \ldots, \tilde{y}^{(B,j)} \sim q_{Y}$. 
Note that \eqref{eq:modified_mine_objective} has the same functional form as the Monte Carlo approximation of the DV representation of $-I(X;Y)$, assuming that the input is uniformly distributed over $\mathcal{X}_t$.
}

\begin{figure}[t]
    \centering
    \includegraphics{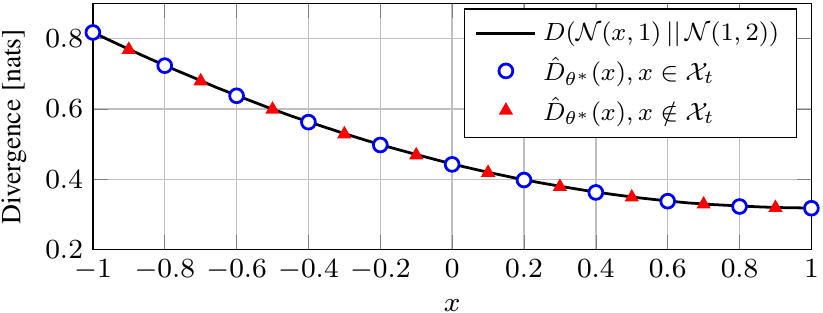}
    \caption{Accuracy and generalization ability of the trained divergence estimator \eqref{eq:modified_mine}, where $\mathcal{X}_t = \{-1, -0.8, \ldots, +1\}$. }
    \label{fig:divergence}
    \vspace*{-0.2cm}
\end{figure}

\emph{Example:} 
Assume that $f_{Y|X=x}$ and $q_Y$ correspond to $\mathcal{N}(x,1)$ and $\mathcal{N}(1,2)$, respectively. 
To optimize the parameters $\theta$, we set $\mathcal{X}_t = \{-1, -0.8, \ldots, +1\}$ and train the statistics network\footnote{The NN architecture and all other training hyperparameters for this example are the same as for case (i) in Tab.~\ref{tab:network_parameters} \NoRev{\Removed{and \ref{tab:hyperparameters}}} below.} using the \NoRev{\Removed{average}} loss \eqref{eq:modified_mine_objective}.
Fig.~\ref{fig:divergence} compares the accuracy of the resulting estimator $\hat{D}_{\theta^*}(x)$ to the true divergence as a function of $x$, where $\theta^*$ refers to the optimized parameters. 
Note that $\hat{D}_{\theta^*}(x)$ generalizes well even to values of $x$ that were not seen during training, as illustrated by the red triangles. \RevC{\Removed{\demo}}
 
\section{Representation of the Reference Distribution}
\label{sec:ndt}

\begin{figure}[t]
    \centering
    \includegraphics{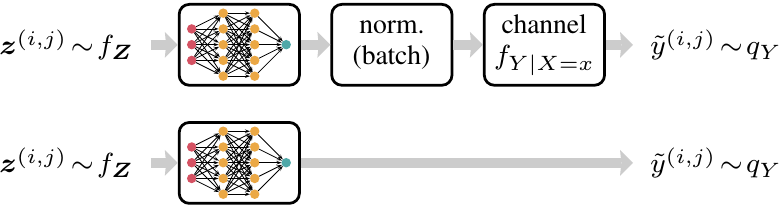}
    \caption{Two approaches for implementing the neural distribution transformer (NDT) that generates samples from the reference distribution $q_Y$. }
    \label{fig:ndt}
    \vspace*{-0.2cm}
\end{figure}

To allow for the gradient-based optimization of the reference distribution $q_Y$, two different approaches are described in the following for generating the samples $\tilde{y}^{(1,j)}, \ldots, \tilde{y}^{(B,j)} \sim q_{Y}$ in \eqref{eq:modified_mine_objective} using NNs. 
The corresponding block diagrams are shown in Fig.~\ref{fig:ndt}. 
Similar to \cite{Aharoni2020}, we refer to the resulting transformation as the neural distribution transformer (NDT).

In the first approach (Fig.~\ref{fig:ndt}, top), the NDT consists of an NN $f_\tau$ with parameters $\tau$, which is then followed by a batch-wise normalization procedure and transmission over the channel. 
\RevC{Thus, this approach generates channel inputs as an intermediate step.}
\RevC{\Removed{In particular, t}T}he normalization procedure enforces the average cost constraint\footnote{Note that this procedure enforces the average cost constraint with equality, even if the inequality constraint was satisfied before the normalization.} via
\begin{align}
    \label{eq:renorm}
    \tilde{s}^{(i,j)} = \frac{{s}^{(i,j)}}{
        c^{-1}\left(\frac{1}{P}\sum_{i=1}^B c({s}^{(i,j)})\right)
    },
\end{align}
where ${s}^{(i,j)} = f_\tau(\vect{z}^{(i,j)})$, $i \in [B]$, $j \in [N_t]$ is the NN output and $\vect{z}^{(i,j)} \in \mathbb{R}^l$ is a random vector sampled from a fixed latent probability distribution $f_{\vect{Z}}$. 
Note that \eqref{eq:renorm} implicitly assumes that the cost function distributes over division, i.e., $c(a/b) = c(a)/c(b)$, which is the case for all cost functions considered in this paper. 
The above approach ensures that $q_Y$ is a valid output distribution for the channel under consideration, given the cost constraint. 
However, it should be noted that it requires a differentiable channel model in order to compute gradients with respect to $\tau$. 

In the second approach (Fig.~\ref{fig:ndt}, bottom), the NDT \RevC{directly generates samples from the channel output alphabet and} simply consists of an NN $f_\tau$ as before but without any additional post-processing, i.e., $\tilde{y}^{(i,j)} = f_\tau(\vect{z}^{(i,j)})$. 
While this approach does not necessarily ensure that $q_Y$ is a valid output distribution for cost-constrained channels (which is not required to obtain an upper bound in \eqref{eq:cup}), it is more universal and can be used even if the channel is only accessible as a black box through input--output samples (e.g., in an experimental setting). 
On the other hand, we found that this representation typically requires more training steps when optimizing the NN parameters $\tau$.

\section{Proposed Algorithm}
\label{sec:algorithm}

\newcommand{\Iouter}{\ensuremath{{M}}}
\newcommand{\Iinit}{\ensuremath{\mathcal{I}_\text{init}}}

The proposed capacity estimation algorithm is detailed in Algorithm~\ref{alg:discrete_upper_bound}. 
It alternates between training the statistics network $T_\theta$ (lines 3--5) and the NDT network $f_\tau$ (lines 6--8) for a total of \RevD{$\Iouter$} iterations. 
The latter optimizes the reference distribution $q_Y$ based on the loss function (cf.~\eqref{eq:cup})
\begin{align}
\begin{split}
    \label{eq:capacity_upper_bound}
    \hat{F}_{\tau}(\gamma) &= \max_{j \in [N_d]} \left[ \hat{D}_\theta(x^{(j)}) - \gamma c(x^{(j)}) \right] \\
    &=\max_{j \in [N_d]} \left[ 
    \frac{1}{B} \sum_{i=1}^{B} T_\theta(y^{(i,j)}, x^{(j)}) \right. \\
  &\RevD{\quad\qquad\quad- \left.\log\left( \frac{1}{B} \sum_{i=1}^{B} e^{T_\theta(\tilde{y}^{(i,j)}, x^{(j)})} \right) - \gamma c(x^{(j)}) \right]},
  \end{split} 
\end{align}
where the dependence of $\hat{F}_\tau(\gamma)$ on the parameters $\tau$ is implicit through the samples $\tilde{y}^{(i,j)}$ generated by the NDT. 
Compared to \eqref{eq:cup}, the outer minimization over $q_Y$ is encapsulated in the NN parameters $\tau$, which are optimized via gradient descent in Algorithm \ref{alg:discrete_upper_bound}. 

The definition of the sets $\mathcal{X}_t$ and $\{x^{(1)}, \ldots, x^{(N_d)}\} = \mathcal{X}_d \subseteq \mathcal{X}$ \NoRev{\Removed{appearing }}in lines 3 and 6 depends on whether the \NoRev{\Removed{channel}} input alphabet is discrete or continuous. 
For channels with discrete input alphabet $\mathcal{X}$, we may set $\mathcal{X}_t = \mathcal{X}_d = \mathcal{X}$.\footnote{In this case, the data generated in line 3 can be reused in line 6.} 
If the input alphabet of the channel is continuous, we assume that the input space has been appropriately discretized and the resulting set of discretized inputs is given by $\mathcal{X}_d$. 
\RevB{\Removed{To ensure that the capacity of the resulting input-discretized channel is close to that of the original channel, one approach is to successively increase the number of discretization points $N_d$ until some form of convergence is reached.}}
A native, but more involved, approach for channels with continuous inputs that does not require any input space discretization is suggested in Sec.~\ref{sec:conclusion}. 

\newcommand{\algtext}[1]{\text{\sf\scriptsize #1}}
\newcommand{\rightcomment}[1]{\tcc*[r]{#1}}
\newcommand\mycommfont[1]{\scriptsize\ttfamily\textcolor{blue}{#1}}
\SetCommentSty{mycommfont}

\setlength{\textfloatsep}{10pt}
\begin{algorithm}[t]
	\small
	\DontPrintSemicolon
	\SetKw{ShortFor}{for}
	\SetKw{KwBreak}{break}
	\SetKw{MyWhile}{while}
	\SetKw{MyIf}{if}
	\SetKw{MySet}{set}
	\SetKw{MyElse}{else}
	\SetKw{MyCompute}{compute}
	\SetKw{KwEach}{each}
	\SetKw{KwAnd}{and}

    \textbf{Inputs:} number of iterations \RevD{$\Iouter$}, batch size $B$, learning rate $\beta$, Lagrange multiplier $\gamma$ (for cost-constrained channels), input sets $\mathcal{X}_t$ (see Sec.~\ref{sec:divergence}) and $\mathcal{X}_d$ (see Sec.~\ref{sec:algorithm})
    
    \smallskip
	
	\For{$l = 1, 2, \dots, \RevD{\Iouter}$}{
        $\forall x^{(j)} \!\in\! \mathcal{X}_t$, $i \!\in\! [B]$: generate $y^{(i,j)} \!\sim\! f_{Y|X=x^{(j)}}, \tilde{y}^{(i,j)} \!\sim\! q_Y$\;
        $L_{\theta} \leftarrow$ compute average loss according to \eqref{eq:modified_mine_objective}\;
        $\theta \leftarrow \theta - \beta \nabla_{\theta} L_{\theta} $ \rightcomment{update statistics network}
        $\forall x^{(j)} \!\in\! \mathcal{X}_d$, $i \!\in\! [B]$: generate $y^{(i,j)} \!\sim\! f_{Y|X=x^{(j)}}, \tilde{y}^{(i,j)} \!\sim\! q_Y$\;
        $\hat{F}_{\tau}(\gamma) \leftarrow$ estimate upper bound according to \eqref{eq:capacity_upper_bound}\;
        $\tau \leftarrow \tau - \beta \nabla_{\tau} \hat{F}_{\tau}(\gamma) $ \rightcomment{update NDT network}
	}
	\Return $\hat{F}_\tau(\gamma)$
	\caption{ {\small Estimation of capacity upper bounds.} }
	\label{alg:discrete_upper_bound}
\end{algorithm}

\section{Numerical Results}

In this section, we numerically evaluate the proposed approach\NoRev{\Removed{by estimating upper bounds on the capacity of various memoryless channels}}.\footnote{The source code to reproduce all numerical results in this paper is available at \url{https://github.com/chaeger/upper_capacity_bounds}.}
Note that for cost-constrained channels, Algorithm \ref{alg:discrete_upper_bound} estimates the capacity in \eqref{eq:cup} as a function of the Lagrange multiplier $\gamma\geq 0$. 
In this case, we use a golden-section search to solve the outer one-dimensional minimization over $\gamma$ in \eqref{eq:cup2}. 

\renewcommand{\arraystretch}{1.0}
\begin{table}[t]
\setlength{\tabcolsep}{0.5em}
\scriptsize
\centering
\vspace{0.15cm}
\caption{NN parameters for the (i) average-power-constrained AWGN, (ii) peak-power-constrained AWGN, (iii) OI, and (iv) NLPN channel.}
\NoRev{%
\begin{tabular}{c|c|ccc|ccc}
\toprule
& & \multicolumn{3}{c}{NDT network $f_\tau$}   & \multicolumn{3}{|c}{statistics~network $T_\theta$} \\
 \midrule
& layer    & input  & hidden& output    & input & hidden     & output  \\ 
\midrule
\multirow{1}{*}{(i)} 
&\# neurons   & $50$ & $2 \times 100$   & $1$ (linear)      & $2$     & $2 \times 100$      & $1$ (linear)          \\ 
\multirow{1}{*}{(ii)}
&\# neurons   & $50$ & $2 \times 100$   & $1$ (tanh)      & $2$     & $2 \times 100$      & $1$ (linear)         \\ 
\multirow{1}{*}{(iii)}
&\# neurons   & $50$ & $2 \times 100$   & $1$ (sigmoid)      & $2$     & $2 \times 100$      & $1$ (linear)          \\ 
\multirow{1}{*}{(iv)}
&\# neurons   & $50$ & $2 \times150$   & $2$ (linear)      & $4$     & $2 \times 150$      & $1$ (linear)          \\ 
\bottomrule
\end{tabular}
}%
\label{tab:network_parameters}
\end{table}

For all considered cases, we use fully-connected NNs with rectified linear unit \NoRev{\Removed{(ReLU)}}activation functions in the hidden layers to represent both the NDT network $f_\tau$ and the statistics network $T_\theta$. 
The NN parameters are summarized in Tab.~\ref{tab:network_parameters}.  
The number of input neurons for $f_\tau$ depends on the latent distribution, which we assume to be a multivariate Gaussian distribution $\mathcal{N}(\vect{0}_l, \vect{I}_l)$ with $l = 50$.
We found that the results are relatively insensitive to the choice of $l$, which mainly affects the initial distribution for $q_Y$ before training. 
Moreover, while we have verified both NDT approaches discussed in Sec.~\ref{sec:ndt}, the following numerical results use the first approach (Fig.~\ref{fig:ndt}, top) since all channel models below are differentiable. 

For the gradient-based optimization steps in Algorithm 1 (lines 5 and 8), we employ the Adam optimizer \cite{Kingma2015} with learning rate \NoRev{$\beta=10^{-3}$} and batch size \NoRev{$B=20000$}. \NoRev{\Removed{ shown in Tab.~\ref{tab:hyperparameters}.}} 
\NoRev{\Removed{Tab.~\ref{tab:hyperparameters} shows the number of iterations and discretization points for each scenario.}}
Lastly, we always pretrain the statistics network $T_\theta$ by running $200$ initial iterations of lines 3--5 in Algorithms~\ref{alg:discrete_upper_bound}, which we found to improve training convergence.

\subsection{AWGN Channel}

\begin{figure}[t]
    \centering
    \includegraphics{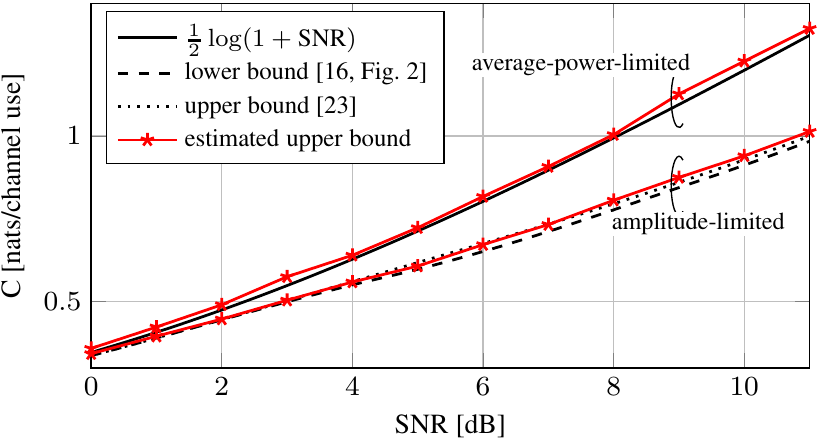}
    \caption{Results for the AWGN channel.}
    \label{fig:results_awgn}
\end{figure}

We start with the additive white Gaussian noise (AWGN) channel $Y = X + Z$, where $X$ is the channel input and $Z \sim \mathcal{N}(0, \sigma^2)$. 
For the average-power-limited case, we have $c(x) = x^2$ as a cost function, in which case $C = \frac{1}{2} \log(1+\SNR)$ with $\SNR = P/\sigma^2$. 
For the amplitude-limited case, we instead have no cost constraint, $|X| \leq A$, and $\text{SNR} = A^2/\sigma^2$. 
In this case, no closed-form analytical capacity expressions exist, but upper and lower bounds have been derived, see, e.g., \cite{McKellips2004},\RevB{\cite{Rassouli2016}}, \cite{Thangaraj2017}\RevB{\Removed{ and references therein}}. 

For the numerical estimation, we set $P=1$ and $A=1$ without loss of generality and vary the SNR by varying $\sigma^2$.
The input space is discretized using $N_d = 15$ uniformly spaced grid points in the intervals $[-2.5, 2.5]$ and $[-1, 1]$ for the average-power-limited and amplitude-limited case, respectively. 
In this paper, we always assume for simplicity that $\mathcal{X}_t = \mathcal{X}_d$, noting that in general the set $\mathcal{X}_d$ can be different from the set $\mathcal{X}_t$ used to train the divergence estimator.
Fig.~\ref{fig:results_awgn} shows the estimated upper bounds \NoRev{after $M=500$ iterations}, where we compare to the \RevB{upper bound in \cite{Rassouli2016} and} lower bound in \cite[Fig.~2]{Thangaraj2017} for the amplitude-limited case. 
Note that the NDT network $f_\tau$ for the amplitude-limited case uses a tanh activation function in the last layer (cf.~Tab.~\ref{tab:network_parameters}) to enforce the amplitude constraint. 
Moreover, due to the absence of a cost constraint, no normalization procedure is applied and the outputs of $f_\tau$ are directly transmitted over the channel to generate the NDT output samples $\tilde{y}^{(i,j)}$. 

\subsection{Optical Intensity Channel}

\begin{figure}[t]
    \centering
    \includegraphics{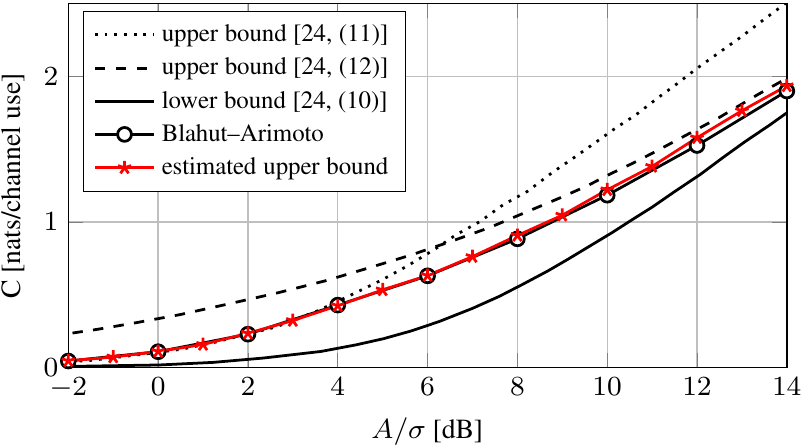}
    \caption{Results for the OI channel with $\alpha=0.4$ (cf.~\cite[Fig.~2]{Lapidoth2009})}
    \label{fig:resuls_oic}
\end{figure}

Next, we consider the optical intensity (OI) channel which is defined by $Y = X + Z$ with $c(x) = x$, $X \in [0, A]$, and $Z \sim \mathcal{N}(0,\sigma^2)$ \cite{Lapidoth2009}. 
We consider the case $\alpha = P/A = 0.4$. 
For the numerical estimation, we set $P = 1$ and discretize the input space using $15$ uniformly spaced grid points in the interval $[0, A]$, where $A = 2.5$.
For this case, the NDT network uses a sigmoid activation (scaled by $A$) in the last layer to ensure that the channel input satisfies the amplitude constraint. 
Results are shown in Fig.~\ref{fig:resuls_oic} \NoRev{after $M=500$ iterations}, where we compare to the upper and lower capacity bounds developed in \cite{Lapidoth2009}, see in particular \cite[Fig.~2]{Lapidoth2009}.
The gap of the estimated upper bound with respect to the lower bound is due to the fact that the latter is not tight. 
Indeed, \RevB{to verify that the channel capacity is close to our estimated upper bound, we used the Blahut--Arimoto algorithm for cost-constrained channels \cite[p.~140]{Csiszar1981} (black circles)}. 
\MinRevB{We also note that tighter analytical upper bounds can potentially be obtained by extending the methodology in \cite{Rassouli2016} to this channel model.} 

\subsection{Nonlinear Phase Noise Channel}
\label{sec:nlpn}

Lastly, we consider an NLPN channel for coherent optical communication which is based on a split-step solution of the nonlinear Schrödinger equation without dispersive effects. 
The resulting complex-valued channel is defined by the recursion
\begin{align}
    \label{eq:nlpn}
    X_{k+1} = X_k e^{\imag \gamma L |X_k|^2/K} + N_{k+1}, \quad  0 \leq k \leq K,
\end{align}
where $X = X_0 \in \mathbb{C}$ is the input, $Y = X_K \in \mathbb{C}$ is the output, $N_{k+1} \sim \mathcal{C}\mathcal{N}(0, \sigma^2/K)$, $\sigma^2$ is the total noise power, $\gamma$ is a nonlinearity parameter, $L$ is the transmission distance, and $c(x) = |x|^2$. 
This channel has a long history in terms of capacity analysis, see \RevB{\cite{Turitsyn2003, Yousefi2011a},}\cite{Shen2018ecoc, Keykhosravi2019}, $\!$\MinRevB{\cite{Reznichenko2020}}\RevB{\Removed{ and references therein}} \MinRevB{and references therein}. 
Here, we use the same parameters as in \cite{Shen2018ecoc, Keykhosravi2019}, i.e., $K=50$, $\sigma^2 = -21.3\,\text{dBm}$, $\gamma = 1.27\,\text{rad}/\text{km}/\text{W}$, and $L = 5000\,$km.

\begin{figure}[t]
    \centering
    \includegraphics{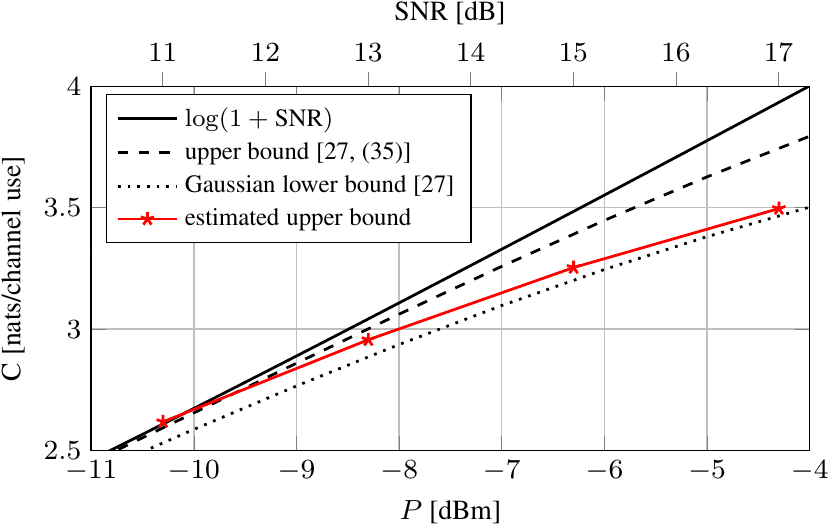}
    \caption{Results for the NLPN channel.}
    \label{fig:resuls_nlpn}
\end{figure}

For the numerical estimation, we consider a renormalized version of \eqref{eq:nlpn}, where $\tilde{X}_k = X_k/\sqrt{P}$. 
The input space of the renormalized channel is discretized using $9$ uniformly spaced grid points in the interval $[-1.75, 1.75]$ for both the real and imaginary part, i.e., $N_d = 81$ total discretization points. 
Separating the real and imaginary parts of the channel input and output, respectively, the number of NDT output neurons is increased to $2$ and the number of input neurons of the statistics network to $4$. 
Fig.~\ref{fig:resuls_nlpn} shows the obtained results \NoRev{after $M=2500$ iterations} as a function of $P$ (see the top axis for a conversion to $\text{SNR} = P/\sigma^2$). 
It can be seen that the estimated upper bound closely follows the lower bound in \cite{Keykhosravi2019}, which is based on a Gaussian input distribution.

\section{Discussion and Future Work}
\label{sec:conclusion}

We have proposed a novel data-driven approach for estimating upper bounds on channel capacity.
\NoRev{\Removed{The approach exploits the dual representation of capacity and does not require knowledge about the underlying channel law.}}
Similar to recent work in \cite{Fritschek2019, Aharoni2020}, the proposed algorithm relies on the \RevC{DV}\RevC{\Removed{Donsker--Varadhan}} representation for estimating divergence, with the consequence that the resulting estimates are neither true upper nor lower bounds \RevB{for a finite sample size \cite{Poole2019}}. 
\RevB{Even assuming an infinite sample size, one cannot guarantee that the resulting estimates are true upper bounds on capacity, which would in principle require a full optimization over the function class $\mathcal{T}$ (cf.~\eqref{eq:donsker}). 
It is therefore important to properly choose the NN and training parameters.  
For example, more training iterations were required for the NLPN channel compared to the other cases to ensure convergence of the NDT and statistics networks.}

An overview of potential alternative divergence estimation approaches can be found in \cite{Fritschek2020} and \cite{Mirkarimi2021comparison}.  
Moreover, \cite{Mirkarimi2021} recently proposed an approach to obtain outer bounds on the achievable rate region of memoryless multiple-access channels by exploiting the upper bounds based on $f$-divergence inequalities from \cite{Sason2016}. 
However, the histogram-based approach to numerically evaluate these bound in \cite{Mirkarimi2021} does not directly lead to a differentiable loss function. 
Therefore, additional modifications (e.g., based on ideas similar to \cite{Ustinova2016}) would be required to be able to use such inequalities in conjunction with the NDT optimization in Algorithm 1. 

Lastly,\NoRev{\Removed{we have assumed that the input alphabet of the channel is either discrete or has been appropriately discretized.}}
\RevB{another reason why the proposed approach does not necessarily compute true upper bounds for continuous-input channels is that the maximization over $x$ is only done approximately via discretization.}
\RevB{To ensure that the capacity of the resulting input-discretized channel is close to that of the original channel, one approach is to successively increase the number of discretization points $N_d$ until convergence.}
For future work, it may be interesting to develop native estimation approaches \NoRev{\Removed{for continuous-input channels }}that do not require such a discretization. 
This could be done, for example, by considering an auxiliary distribution over the input space and casting the maximization in \eqref{eq:cup} as an optimization problem over this auxiliary distribution. 
Similar to the NDT, the auxiliary distribution could then again be parameterized using an NN.

\end{document}